\documentclass[english,oneside,12pt]{elsart}
\usepackage[T1]{fontenc}
\usepackage[latin1]{inputenc}
\usepackage{graphicx}
\usepackage{amssymb}

\makeatletter

\providecommand{\tabularnewline}{\\}



\usepackage{babel}
\makeatother

\begin{document}
\begin{frontmatter}

\title{A Fortran 90 program to solve the Hartree-Fock equations for interacting
spin-$\frac{1}{2}$ Fermions confined in Harmonic potentials}

\author{Hridis Kumar Pal$^{1,2}$, Alok Shukla$^{3}$}

\address{Physics Department, Indian Institute of Technology, Powai, Mumbai
400076, INDIA}

\thanks{Done in partial fulfillment of the requirements for the degree of
Master of Science in Physics at the Indian Institute of Technology,
Bombay, India. }

\thanks{Present address: Department of Physics, University of Florida, Gainesville,
FL 32611-8440, USA. email:hridispal@phys.ufl.edu}

\thanks{Author to whom all the correspondence should be addressed. email:shukla@phy.iitb.ac.in}

\begin{abstract}
A set of weakly interacting spin-$\frac{1}{2}$ Fermions, confined
by a harmonic oscillator potential, and interacting with each other
via a contact potential, is a model system which closely represents
the physics of a dilute gas of two-component Fermionic atoms confined
in a magneto-optic trap. In the present work, our aim is to present
a Fortran 90 computer program which, using a basis set expansion technique,
solves the Hartree-Fock (HF) equations for spin-$\frac{1}{2}$ Fermions
confined by a three-dimensional harmonic oscillator potential, and
interacting with each other via pair-wise delta-function potentials.
Additionally, the program can also account for those anharmonic potentials
which can be expressed as a polynomial in the position operators $x,$
$y$, and $z$. Both the restricted-HF (RHF), and the unrestricted-HF
(UHF) equations can be solved for a given number of Fermions, with
either repulsive or attractive interactions among them. The option
of UHF solutions for such systems also allows us to study possible
magnetic properties of the physics of two-component confined atomic
Fermi gases, with imbalanced populations. Using our code we also demonstrate
that such a system exhibits shell structure, and follows Hund's rule.
\end{abstract}
\end{frontmatter}
\textbf{Program Summary} \\
 \emph{Title of program:} trap.x \\
 \emph{Catalogue Identifier:} \\
 \emph{Program summary URL:} \\
 \emph{Program obtainable from:} CPC Program Library, Queen's University
of Belfast, N. Ireland \\
 \emph{Distribution format:} tar.gz\\
 \emph{Computers :} PC's/Linux, Sun Ultra 10/Solaris, HP Alpha/Tru64,
IBM/AIX\\
 \emph{Programming language used:} mostly Fortran 90\\
 \emph{Number of bytes in distributed program, including test data,
etc.:} size of the gzipped tar file 371074 bytes\\
 \emph{Card punching code:} ASCII\\
 \emph{Nature of physical problem:} The simplest description of a
spin $\frac{1}{2}$ trapped system at the mean field level is given
by the Hartree-Fock method. This program presents an efficient approach
of solving these equations. Additionally, this program can solve for
time-independent Gross-Pitaevskii and Hartree-Fock equations for bosonic
atoms confined in a harmonic trap. Thus the combined program can handle
mean-field equations for both the fermi and the bose particles. \\
 \emph{Method of Solution:} The solutions of the Hartree-Fock equation
corresponding to the fermi systems in atomic traps are expanded as
linear combinations of simple-harmonic oscillator eigenfunctions.
Thus, the Hartree-Fock equations which comprises of a set of nonlinear
integro-differential equation, is transformed into a matrix eigenvalue
problem. Thereby, its solutions are obtained in a self-consistent
manner, using methods of computational linear algebra.\\
 \emph{Unusual features of the program:} None

\section{Introduction}

Over the last several years, there has been an enormous amount of
interest in the physics of dilute Fermi gases confined in magneto-optic
traps\cite{fermi-gas-exp,new-exp,fermi-theory-bec-bcs,ogren-shell,fermi-gas-theory-phases}.
With the possibility of tuning the atomic scattering lengths from
the repulsive regime to an attractive one using the Feshbach resonance
technique, there has been considerable experimental activity in looking
for phenomenon such as superfluidity, and other phase transitions
in these systems\cite{fermi-gas-exp,new-exp}. This has led to equally
vigorous theoretical activity starting from the studies of so-called
BEC-BCS crossover physics\cite{fermi-theory-bec-bcs}, search for
shell-structure in these systems\cite{ogren-shell}, to the study
of more complex phases\cite{fermi-gas-theory-phases}. As far as the
spin of the fermions is concerned, most attention has been given to
the cases of two-component gases which can be mapped to a system of
spin-$\frac{1}{2}$ atoms\cite{fermi-theory-bec-bcs,ogren-shell}.
Therefore, in our opinion, a quantum-mechanical study of spin-$\frac{1}{2}$
fermions moving in a harmonic oscillator potential, and interacting
via a pair-wise delta function potential, can help us achieve insights
into the physics of dilute gases of trapped fermionic atoms.

With the aforesaid aims in mind, the purpose of this paper is to describe
a Fortran 90 computer program developed by us which can solve the
Hartree-Fock equations for spin-$\frac{1}{2}$ fermions moving in
a three-dimensional (3D) harmonic oscillator potential, and interacting
via delta-function potential. A basis set approach has been utilized
in the program, in which the single-particle orbitals are expanded
as a linear combination of the 3D simple harmonic oscillator basis
functions, expressed in terms of Cartesian coordinates. The program
can solve both the restricted-Hartree-Fock (RHF), and the unrestricted
Hartree-Fock (UHF) equations, the latter being useful for  fermi gases
with imbalanced populations. We would like to clarify, that as far
as the applications of this approach to dilute Fermi gases is concerned,
at present it is not possible to reach the thermodynamic limit of
very large $N$, where $N$ is the total number of atoms in the trap.
However, we believe that by solving the HF equations for a few tens
of atoms, one may be able to achieve insights into the microscopic
aspects such as the nature of pairing in such systems. This program
is an extension of an earlier program developed in our group, aimed
at solving the time-independent Gross-Pitaevskii equation (GPE) for
harmonically trapped Bose gases\cite{shukla}. Thus the combined total
program accompanying this paper can now solve for both Bose and Fermi
systems, confined to move in a harmonic oscillator potential, with
mutual interactions of the delta-function form. As with our earlier
boson program, because of the use of a Cartesian harmonic oscillator
basis set, the new program can handle trap geometries ranging from
spherical to completely anisotropic, and it can also account for those
trap anharmonicities which can be expressed as polynomials in the
Cartesian coordinates. The nature of interparticle interactions, i.e.,
whether they are attractive or repulsive, also imposes no restrictions
on the program. We note that Yu \emph{et al.}\cite{ogren-shell}\emph{
}have recently described a Hartree-Fock approach for dealing with
two-component fermions confined in harmonic traps with spherical symmetry,
employing a finite-difference-based numerical approach. However, we
would like to emphasize that, as mentioned earlier, our approach is
more general in that it is not restricted to any particular trap symmetry.
Apart from describing the program, we also present and discuss several
of its applications. With the aim of exploring the shell-structure
in trapped fermionic atoms, using our UHF approach we compute the
addition energy for spherically trapped fermions for various particle
numbers, and obtain results consistent with a shell-structure and
Hund's rule.

The remainder of the paper is organized as follows. In the next section
we discuss the basic theoretical aspects of our approach. In section
\ref{sec-program}, we briefly describe the most important subroutines
that comprise the new enlarged program. Section \ref{sec:install}
contains a brief note on how to install the program and prepare the
input files. In section \ref{sec-results} we discuss results of several
example runs of our program for different geometries. In the same
section, we also discuss issues related to the convergence of the
procedure. Finally, in section \ref{sec-conclusions}, we end this
paper with a few concluding remarks.

\section{Theory}

\label{sec-theory}

We consider a system of $N$ identical spin-$\frac{1}{2}$ particles
of mass $m$, moving in a 3D potential with harmonic and anharmonic
terms, interacting with each other via a pair-wise delta function
potential. The Hamiltonian for such a system can be written as \begin{equation}
H=\sum_{i=1}^{N}h({\bf r}_{i})+g\sum_{i>j}^{N}\delta({\bf r}_{i}-{\bf r}_{j}),\label{eq:ham}\end{equation}

where ${\bf r}_{i}$ represents the position vector of $i-$th particle,
$g$ represents the strength of the delta-function interaction, and
$h({\bf r}_{i})$ denotes the one-particle terms of the Hamiltonian\begin{equation}
h({\bf r}_{i})=-\frac{\hbar^{2}}{2m}\mathbf{\nabla}_{i}^{2}+\frac{1}{2}m(\omega_{x}^{2}x_{i}^{2}+\omega_{y}^{2}y_{i}^{2}+\omega_{z}^{2}z_{i}^{2})+V^{anh}(x_{i},y_{i},z_{i}),\label{eq:h1}\end{equation}
where $\omega_{x}$, $\omega_{y}$ and $\omega_{z}$ are the angular
frequencies of the external harmonic potential in the $x$, $y$ and
$z$ directions, respectively, and $V^{anh}(x_{i},y_{i},z_{i})$ represents
any anharmonicity in the potential. In order to parametrize the strength
of the delta-function interactions, we use the formula $g=\frac{4\pi\hbar^{2}a}{m}$
in our program, where $a$ is the $s$-wave scattering length for
the atoms. Next we will obtain the RHF and the UHF equations for the
system. 

Assuming that $N=2n$, and that the many-particle wave function of
the system can be represented by a single closed-shell Slater determinant,
the RHF equations for the $n$ doubly occupied orbitals $\{\psi_{i}({\bf r}),\: i=1,\ldots,n\}$
of the system are obtained to be\cite{szabo}\begin{equation}
(h+g\sum_{j=1}^{n}|\psi_{j}({\bf r})|^{2})\psi_{i}({\bf r})=\epsilon_{i}\psi_{i}({\bf r}).\label{eq:rhf}\end{equation}

Similarly, for a system with $n_{1}$ up-spin ($\alpha$) fermions,
and $n_{2}$ down-spin ($\beta$) fermions $(n_{1}+n_{2}=N)$, the
UHF equations for the up-spin orbitals can be written as\cite{szabo}\begin{equation}
(h+g\sum_{j=1}^{n_{2}}|\psi_{j}^{(\beta)}({\bf r})|^{2})\psi_{i}^{(\alpha)}({\bf r})=\epsilon_{i}\psi_{i}^{(\alpha)}({\bf r}),\label{eq:uhf}\end{equation}
where $\{\psi_{i}^{(\alpha)}({\bf r}),\: i=1,\ldots,n_{1}\}$ and
$\{\psi_{j}^{(\beta)}({\bf r}),\: j=1,\ldots,n_{2}\}$, represent
the occupied orbitals corresponding to the up and the down spins,
respectively. Similar to the the UHF equations for the down-spin orbitals
can be deduced easily from Eq. (\ref{eq:uhf}). As in our earlier
work on the bosonic systems\cite{shukla}, we adopt a basis-set approach
and expand the HF orbitals in terms of the 3D Harmonic oscillator
basis functions. This approach is fairly standard, and is well-known
as the Hartree-Fock-Roothan procedure in the quantum chemistry community\cite{szabo}.
Thus, for the RHF case, the orbitals are expressed as\begin{equation}
\psi_{i}({\bf r})=\sum_{j=1}^{N_{basis}}C_{ji}\Phi_{n_{xj},n_{yj},n_{zj}}({\bf r})=\sum_{j=1}^{N_{basis}}C_{ji}\phi_{n_{xj}}(x)\phi_{n_{yj}}(y)\phi_{n_{zj}}(z),\label{eq:basis-rhf}\end{equation}

where $C_{ji}$ represents the coefficient corresponding to the $j$-th
3D harmonic oscillator basis function $\Phi_{n_{xj},n_{yj},n_{zj}}({\bf r})$,
in the expansion of the $i$-th occupied orbital $\psi_{i}({\bf r})$,
and $N_{basis}$ is the total number of basis functions used. Note
that $\Phi_{n_{xj},n_{yj},n_{zj}}({\bf r})$ is itself a product of
three linear harmonic oscillator eigenfunctions of quantum numbers
$n_{xj},$ $n_{yj}$, and $n_{zj}$. Therefore, a set of functions
$\Phi_{n_{xj},n_{yj},n_{zj}}({\bf r})$, for different values of $n_{xj},$
$n_{yj}$, and $n_{zj}$, will constitute an orthonormal basis set,
leading to an overlap matrix which is identity matrix. For the UHF
case, the corresponding expansion for up-spin particles is\begin{equation}
\psi_{i}^{(\alpha)}({\bf r})=\sum_{j=1}^{N_{basis}}C_{ji}^{(\alpha)}\phi_{n_{xj}}(x)\phi_{n_{yj}}(y)\phi_{n_{zj}}(z),\label{eq:basis-uhf}\end{equation}
 from which the expansion for the down-spin particles can be easily
deduced. Upon substituting Eqs. (\ref{eq:basis-rhf}) and (\ref{eq:basis-uhf}),
in Eqs. (\ref{eq:rhf}) and (\ref{eq:uhf}), respectively, one can
obtain the matrix forms of the RHF/UHF equations\cite{szabo}. As
outlined in our earlier work\cite{shukla}, numerical implementation
of the approach is carried out in the so-called harmonic oscillator
units, in which the unit of length is the quantity $a_{x}=\sqrt{\frac{\hbar}{m\omega_{x}}}$
, and that of energy is $\hbar\omega_{x}$. The resulting matrix equation
for the RHF case is \begin{equation}
\hat{F}\hat{C}_{(i)}=\tilde{\epsilon_{i}}\hat{C}_{(i)},\label{eq-eigval-rhf}\end{equation}
 where $\hat{C}_{(i)}$ represents the column vector containing expansion
coefficients $\{C_{ji},\: j=1,\ldots,N_{basis}\}$ of $\psi_{i}$,
$\tilde{\epsilon_{i}}$ is the corresponding energy eigenvalue, and
the elements of the Fock matrix $\hat{F}$ are given by\begin{equation}
\hat{F_{i,j}}=E_{i}\delta_{i,j}+V_{i,j}^{anh}+g\sum_{k,l=1}^{N_{basis}}\tilde{J}_{i,j,k,l}D_{k,l}.\label{eq-fock-rhf}\end{equation}
Above\begin{equation}
E_{i}=(n_{xi}+\frac{1}{2})+(n_{yi}+\frac{1}{2})\gamma_{y}+(n_{zi}+\frac{1}{2})\gamma_{z},\label{eq-ediag}\end{equation}
 expressed in terms of aspect ratios $\gamma_{y}=\frac{\omega_{y}}{\omega_{x}}$
and $\gamma_{z}=\frac{\omega_{z}}{\omega_{x}}$, $V_{i,j}^{anh}$
are the matrix elements of the anharmonic term in the confining potential,
$D_{k,l}=\sum_{i=1}^{n}C_{ki}C_{li}$ is a density-matrix element,
and $\tilde{J}_{i,j,k,l}$ represents the 3D two-fermion repulsion
matrix defined as\begin{equation}
\tilde{J}_{i,j,k,l}=J_{n_{xi}n_{xj}n_{xk}n_{xl}}J_{n_{yi}n_{yj}n_{yk}n_{yl}}J_{n_{zi}n_{zj}n_{zk}n_{zl}.}\label{eq-jtilde}\end{equation}
Each one of the $J$ matrices in Eq. (\ref{eq-jtilde}), corresponding
to the three Cartesian directions, can be written in the form\begin{equation}
J_{n_{i}n_{j}n_{k}n_{l}}=\int_{-\infty}^{\infty}d\xi\phi_{n_{l}}(\xi)\phi_{n_{k}}(\xi)\phi_{n_{j}}(\xi)\phi_{n_{i}}(\xi),\label{eq-jmat}\end{equation}
where $\xi$ is the corresponding Cartesian coordinate in the harmonic
oscillator units. An analytical expression for $J_{n_{i}n_{j}n_{k}n_{l}}$
can be found in our earlier work\cite{shukla}. In the UHF case, one
obtains two matrix equations for the up/down-spin particles of the
form\begin{equation}
\hat{F}^{(\alpha)}\hat{C}_{(i)}^{(\alpha)}=\tilde{\epsilon}_{i}^{(\alpha)}\hat{C}_{(i)}^{(\alpha)},\label{eq-eigval-uhf}\end{equation}

where the $\hat{F}^{(\alpha)}$ represents the Fock matrix for the
up-spin particles given by $\hat{F_{i,j}}^{(\alpha)}=E_{i}\delta_{i,j}+V_{i,j}^{anh}+g\sum_{k,l=1}^{N_{basis}}\tilde{J}_{i,j,k,l}D_{k,l}^{(\beta)}$,
$\tilde{\epsilon}_{i}^{(\alpha)}$ is the energy eigenvalue, and $D_{k,l}^{(\beta)}=\sum_{i=1}^{n_{2}}C_{ki}^{(\beta)}C_{li}^{(\beta)}$,
are the elements of the down-spin density matrix. We can easily deduce
the form of the Fock equation for the down-spin particles from Eq.
(\ref{eq-eigval-uhf}). In our program, HF Eqs. (\ref{eq-eigval-rhf})
and (\ref{eq-eigval-uhf}) are solved employing the self-consistent
field (SCF) procedure, which requires the iterative diagonalization
of the Fock equations\cite{szabo}.

\section{Description of the program}

\label{sec-program}

In this section we briefly describe the main program and various subroutines
which constitute the entire module. As mentioned in the Introduction,
the present program is an extension of our earlier program for bosons\cite{shukla}.
Thus the new program, which compiles as \emph{trap.x}, can solve for:
(a) time-independent Gross-Pitaevskii equation for bosons, and (b)
Hartree-Fock equations for fermions, confined in a trap. Therefore,
most of the changes in the present program, as compared the earlier
bosonic program, are related to its added fermionic HF capabilities.
However, we have also tried to optimize the earlier bosonic module
of the program wherever possible. A README file associated with this
program lists all its subroutines. Thus, in what follows, we will
describe only those subroutines which are either new (fermion related),
or modified, as compared to the older bosonic code\cite{shukla}.
For an account of the older subroutines not described here, we refer
the reader to our earlier work\cite{shukla}. Additionally, with the
aim of making the calculations faster, in the present code, we use
the diagonalization routines of LAPACK library\cite{lapack}, which
requires the linking of our code to that library. Therefore, for this
program to work, the user must have the LAPACK/BLAS program libraries
installed on his/her computer system. The letter F or B has been included
in parenthesis after the name of each subroutine to show whether the
subroutine is useful for Fermionic or Bosonic calculations. If it
is applicable for both, we denote this by writing BF.

\subsection{Main Program OSCL (BF)}

This is the main program of our package which reads the input data,
dynamically allocates relevant arrays, and then calls other subroutines
to perform tasks related to the remainder of the calculations. The
main modification in this program, as compared to its earlier version\cite{shukla},
is that it now allows for input related to fermionic HF calculations.
Thus, the user now has to specify whether the particles considered
are bosons or fermions. If the particles considered are fermions,
one has to further specify whether the RHF or the UHF calculations
are desired. For the case of UHF calculations, the user also needs
to specify the number of up- and down-spin orbitals. Because of the
dynamic array allocation throughout, no data as to the size of the
arrays is needed from the user. The program will stop only if it exhausts
all the available memory on the computer. There is one major departure
in the storage philosophy in the present version of the code as compared
to the previous one\cite{shukla} in that now only the lower/upper
triangles of most of the real-symmetric matrices (such as the Fock
matrix) are stored in the linear arrays in the packed format. This
not only reduces the memory requirements roughly by a factor of two,
but also leads to faster execution of the code.

\subsection{BECFERMI\_DRV (BF)}

This is the modified version of the old subroutine BEC\_DRV, and is
called from the main program OSCL. As its name suggests, it is the
driver routine for performing: (a) calculations of the bose condensate
wave function for bosons, or (b) solving the RHF/UHF equations for
fermions. Apart from allocating a few arrays, the main task of this
routine is to call either: (a) routines BOSE\_SCF or BOSE\_STEEP depending
upon whether the user wants to use the SCF or the steepest-descent
approach meant for solving the GPE\cite{shukla}, or (b) routines
FERMI\_RHF or FERMI\_UHF depending on whether the RHF or UHF calculations
are to be performed.

\subsection{FERMI\_RHF (F)}

This subroutine solves the RHF equations for the fermions in a trap
using the SCF procedure, mentioned earlier. Its main tasks are as
follows:

\begin{enumerate}
\item Allocate various arrays needed for the SCF calculations 
\item Setup the starting orbitals. This is achieved by diagonalizing the
one-particle part of the Hamiltonian. 
\item Perform the SCF calculations. For this purpose, the two-particle integrals
$J_{i,j,k,l}$ (cf. Eq. (\ref{eq-jtilde})) are calculated during
each iteration\cite{shukla}. If the user has opted for Fock matrix/orbital
mixing, it is implemented using the formula \[
R^{(i)}=xmix\: R^{(i)}+(1-xmix)\: R^{(i-1)},\]
where $R^{(i)}$ is the quantity under consideration in the $i$-th
iteration, and parameter $xmix$ quantifying the mixing is user specified.
Thus, if Fock matrix mixing has been opted, $xmix$ specifies the
fraction of the new Fock matrix in the total Fock matrix in the $i$-th
iteration. If the user has opted for the orbital mixing, then each
occupied orbital is mixed as per the formula above. The Fock matrix
constructed in each iteration is diagonalized using the LAPACK routine
DSPEVX\cite{lapack}, which can obtain a selected number of eigenvalues/eigenvectors
of a real-symmetric matrix, as against traditional diagonalizers which
calculate the entire spectrum of such matrices. We use DSPEVX during
the SCF iterations to obtain only the occupied orbitals and their
energies, thereby, leading to a much faster completion of the SCF
process in comparison to using a diagonalizer which computes all the
eigenvalues/vectors of the Fock matrix. The occupied orbitals are
identified according to the \emph{aufbau} principle.
\item The total energy and the wave function obtained after every iteration
are written in various data files so that the progress of the calculation
can be monitored. This process continues until the required precision
(user specified) in the total HF energy is obtained. 
\end{enumerate}

\subsection{FERMI\_UHF (F)}

In structure and philosophy this subroutine is similar to FERMI\_RHF,
except that its purpose is to solve the UHF equations for interacting
spin-$\frac{1}{2}$ fermions confined in a harmonic potential. Because
there are two separate Fock equations corresponding to the up- and
the down-spin fermions, the computational effort associated with this
subroutine is roughly twice that of routine FERMI\_RHF.

\subsection{BOSE\_SCF (B)}

This subroutine aims at solving the time-independent GPE for bosons
using the iterative diagonalization approach, and was described in
our earlier paper\cite{shukla}. The diagonalizing routine which was
being used for the purpose obtained all the eigenvalues and eigenvectors
of the GPE, which is quite time consuming for calculations involving
large basis sets. Since the condensate corresponds to the lowest-energy
solution of the GPE, using diagonalizing routines which obtain all
its eigenvalues and eigenvectors is wasteful. Therefore, in the new
version of BOSE\_SCF we now use the LAPACK\cite{lapack} routine DSPEVX
to obtain the lowest eigenvalue and the eigenvector of the Hamiltonian
during the SCF cycles, leading to substantial improvements in speed.

\subsection{BOSE\_STEEP (B)}

This subroutine aims at solving the time-independent GPE for bosons
using the steepest-descent method, and was also described in our earlier
paper\cite{shukla}. In this routine, the main computational step
is multiplication of a trial vector by the matrix representation of
the Hamiltonian. In the earlier version of the code, because the entire
Hamiltonian was being stored in a two-dimensional array, we used the
Fortran 90 intrinsic subroutine MATMUL for the purpose. However, now
that we only store the upper triangle of the Hamiltonian in a linear
array, it is fruitful to use an algorithm which utilizes this aspect.
Therefore, we have replaced the call to MATMUL by a call to a routine
called MATMUL\_UT written by us. This has also lead to significant
speed improvements.

\subsection{MATMUL\_UT (B)}

As mentioned in the previous section, the aim of this subroutine is
to multiply a vector by a real-symmetric matrix, whose upper triangle
is stored in a linear array. This routine is called from the subroutine
BOSE\_STEEP, and it utilizes a straightforward algorithm for achieving
its goals by calling two BLAS\cite{lapack} functions DDOT and DAXPY.

\subsection{Plotting Subroutines (BF)}

We have also significantly improved the capabilities of the program
as far as plotting of the orbitals and the associated densities is
concerned. Now the orbitals, or corresponding densities, can be computed
both on one-dimensional and two-dimensional spatial grids, along user-specified
directions, or planes. The driver subroutine for the purpose is called
PLOT\_DRV, which in turn calls the specific subroutines suited for
the calculations. These subroutines are PLOT\_1D, and PLOT\_1D\_UHF
for the one-dimensional plots, and PLOT\_2D and PLOT\_2D\_UHF for
the planar plots. The output of this module is written in a file called
\texttt{orb\_plot.dat}, which can be directly used in plotting programs
such as \texttt{gnuplot }or \texttt{xmgrace}.

\section{Installation, input files, output files}

\label{sec:install}

In our earlier paper, we had described in detail how to install, compile,
and run our program on various computer systems\cite{shukla}. Additionally,
we had explained in a step-by-step manner how to prepare the input
file meant for running the code, and also the contents of a typical
output file\cite{shukla}. Because, various aspects associated with
the installation and running of the program remain unchanged, except
for some minor details, we prefer not to repeat the same discussion.
Instead, we refer the reader to the README file in connection with
various details related to the installation and execution of the program.
Additionally, the file '\texttt{input\_prep.pdf}' explains how to
prepare a sample input file. Several sample input and output files
corresponding to various example runs are also provided with the package.

\section{Calculations and Results}

\label{sec-results}

In this section we report results of some of the calculations performed
by our code on fermionic systems. We present both RHF and UHF calculations
for various types of traps. Further, we discuss some relevant issues
related to the convergence of the calculations.

\subsection{RHF Calculations: total energy convergence}

In this section our aim is to investigate the convergence properties
of the total HF energy of our program with respect to: (a) number
of particles in the trap, (b) symmetry of the confining potential,
(c) nature and strength of interactions, and (d) number of basis functions
employed in the calculations. As far as the number of particles is
concerned, we have considered two closed-shell systems namely with
two particles ($N=2$), and with eight particles ($N=8$). For $N=2$
case, calculations have been performed for all possible trap geometries
ranging from a spherical trap to a completely anisotropic trap. During
these calculations, we have considered both attractive and repulsive
interactions, corresponding to negative and positive scattering lengths,
respectively. The magnitude of the scattering length ($|a|$) employed
in these calculations ranges from $0.01a_{x}$ to $0.8a_{x}$. To
put these numbers in perspective, we recall that in most of the atomic
traps, $a_{x}\approx1.0$ $\mu$m, and for a two-component $^{6}$Li
trapped gas, the estimated value of the scattering length is anomalously
large $a\approx-2160a_{0}$\cite{scatt-length-li6}, where $a_{0}$
is the Bohr radius. Thus, for this very strongly interacting system,
the scattering length $a\approx-0.11a_{x}$, is well within the range
of the scattering lengths considered in these calculations. Therefore,
the systems considered here---ranging from weakly interacting ones
to very strongly interacting ones---truly test our numerical methods. 

\begin{table}
\begin{tabular}{|c|c|c|c|c|c|}
\hline 
 &  & $a=0.1a_{x}$ & $a=0.2a_{x}$ & $a=0.4a_{x}$ & $a=0.8a_{x}$\tabularnewline
\hline
\hline 
$nmax$ & $N_{basis}$ & $E_{HF}$ & $E_{HF}$ & $E_{HF}$ & $E_{HF}$\tabularnewline
\hline
\hline 
$2$ & $10$ & $3.150676$ & $3.285944$ & $3.521934$ & $3.906583$\tabularnewline
\hline 
$4$ & $35$ & $3.149708$ & $3.283415$ & $3.517600$ & $3.904041$\tabularnewline
\hline 
$6$ & $84$ & $3.149568$ & $3.283145$ & $3.517392$ & $3.904014$\tabularnewline
\hline 
$8$ & $165$ & $3.149546$ & $3.283118$ & $3.517390$ & $3.903925$\tabularnewline
\hline 
$10$ & $286$ & $3.149543$ & $3.283117$ & $3.517388$ & $3.903892$\tabularnewline
\hline
$12$ & $455$ & $3.149543$ & $3.283117$ & $3.517386$ & $3.903886$\tabularnewline
\hline
$14$ & $680$ & $3.149543$ & $3.283117$ & $3.517385$ & $3.903885$\tabularnewline
\hline
\end{tabular}

\caption{Convergence of total HF energy ($E_{HF}$) for a spherically symmetric
trap containing two particles, with respect to the size of the basis
set, for various positive values (repulsive interactions) of the scattering
length. Above, $nmax$ is the maximum value of the quantum number
of the SHO basis function in a given direction, and $N_{basis}$ is
the total number of basis functions corresponding to a given value
of $nmax$. In some cases, Fock matrix mixing approach was used to
achieve convergence.}

\label{tab:hf-iso-rep}
\end{table}

The results of our calculations are presented in tables \ref{tab:hf-iso-rep}---\ref{tab:hf-8part}.
For $N=2$ system, we performed these calculations in order to understand
the convergence behavior of the total energy with respect to the basis
set size, with the goal of a high precision (six decimal digit convergence)
in the total energy. Such high accuracy on larger systems will be
computationally much more expensive, and, therefore, our aim behind
the study of $N=8$ system was to understand the role of number of
particles on our results. The next larger closed-shell system will
correspond to $N=20$, but we have not studied that here, because,
in our opinion, such calculations will not lead to any newer insights
into our approach. Next we discuss our results on these systems individually. 

With the aim of a more detailed exposition of the convergence behavior
for repulsive and attractive interactions, for $N=2$ system corresponding
to an isotropic trap, we present our results for the positive and
negative scattering lengths in separate tables \ref{tab:hf-iso-rep}
and \ref{tab:hf-iso-attr}. For the rest of the cases, results for
the attractive and the repulsive interactions are presented in the
same tables. Upon examining our results for $N=2$ case (\emph{cf.}
tables \ref{tab:hf-iso-rep}---\ref{tab:hf-aniso}), we conclude that
for the case of repulsive interactions, calculations always exhibit
convergence from above on $E_{HF}$, with respect to the basis set
size. In order to achieve six-digit accuracy for repulsive interactions,
one needs to use relatively large basis sets, although a three-digit
accuracy can be obtained using considerably smaller basis sets. However,
quite expectedly, a drastically distinct convergence behavior is seen
for the cases involving attractive interactions. It is obvious that
for the attractive interactions, for sufficiently large scattering
length, the HF method will not be applicable, and will exhibit instabilities
because of pair formation. For relatively weaker attractive interactions,
one again encounters convergence from above, as was the case for repulsive
interactions. But, as the strength of the attractive interactions
increases, the convergence with respect to the basis set size becomes
more difficult to achieve, and for $|a|>0.3a_{x}$ (with $a<0$),
this property is completely lost, and the HF method begins to exhibit
unstable behavior. %
\begin{table}
\begin{tabular}{|c|c|c|c|c|c|}
\hline 
 &  & $a=-0.1a_{x}$ & $a=-0.2a_{x}$ & $a=-0.3a_{x}$ & $a=-0.4a_{x}$\tabularnewline
\hline
\hline 
$nmax$ & $N_{basis}$ & $E_{HF}$ & $E_{HF}$ & $E_{HF}$ & $E_{HF}$\tabularnewline
\hline
\hline 
$2$ & $10$ & $2.830199$ & $2.637330$ & $2.418234$ & $2.171854$\tabularnewline
\hline 
$4$ & $35$ & $2.827878$ & $2.622865$ & $2.368604$ & $2.045741$\tabularnewline
\hline 
$6$ & $84$ & $2.827266$ & $2.617430$ & $2.340043$ & $1.935626$\tabularnewline
\hline 
$8$ & $165$ & $2.827091$ & $2.615260$ & $2.321750$ & $1.816197$\tabularnewline
\hline 
$10$ & $286$ & $2.827038$ & $2.614355$ & $2.309079$ & $1.662494$\tabularnewline
\hline
$12$ & $455$ & $2.827021$ & $2.613963$ & $2.299711$ & $1.448111$\tabularnewline
\hline
$14$ & $680$ & $2.827016$ & $2.613787$ & $2.292337$ & $1.115306$\tabularnewline
\hline
$16$ & $969$ & $2.827014$ & $2.613706$ & $2.286104$ & $0.767242$\tabularnewline
\hline
$18$ & $1330$ & $2.827013$ & $2.613667$ & $2.280232$ & $0.287303$\tabularnewline
\hline
\end{tabular}

\caption{Convergence of total HF energy for a spherically symmetric trap containing
two particles, with respect to the size of the basis set, for various
negative values (attractive interactions) of the scattering length
$a$. Various symbols have the same meaning as in table \ref{tab:hf-iso-rep}. }

\label{tab:hf-iso-attr}
\end{table}

\begin{table}
\begin{tabular}{|c|c|c|c|c|c|c|c|}
\hline 
 &  &  & $a=-0.3a_{x}$ & $a=-0.1a_{x}$ & $a=0.1a_{x}$ & $a=0.2a_{x}$ & $a=0.4a_{x}$\tabularnewline
\hline
\hline 
$nxmax$ & $nzmax$ & $N_{basis}$ & $E_{HF}$ & $E_{HF}$ & $E_{HF}$ & $E_{HF}$ & $E_{HF}$\tabularnewline
\hline
\hline 
$2$ & $0$ & $6$ & $3.843213$ & $4.540956$ & $5.080053$ & $5.303210$ & $5.689668$\tabularnewline
\hline 
$2$ & $2$ & $18$ & $3.776326$ & $4.536093$ & $5.077093$ & $5.293658$ & $5.662586$\tabularnewline
\hline 
$4$ & $0$ & $15$ & $3.757823$ & $4.536934$ & $5.078865$ & $5.300842$ & $5.687999$\tabularnewline
\hline 
$4$ & $2$ & $45$ & $3.652833$ & $4.531485$ & $5.075905$ & $5.291342$ & $5.660857$\tabularnewline
\hline 
$4$ & $4$ & $75$ & $3.627775$ & $4.530813$ & $5.075670$ & $5.290704$ & $5.659490$\tabularnewline
\hline
$6$ & $2$ & $84$ & $3.566045$ & $4.530250$ & $5.075782$ & $5.291219$ & $5.660855$\tabularnewline
\hline
$6$ & $4$ & $140$ & $3.524613$ & $4.529542$ & $5.075548$ & $5.290582$ & $5.659487$\tabularnewline
\hline
$6$ & $6$ & $196$ & $3.510247$ & $4.529419$ & $5.075520$ & $5.290518$ & $5.659382$\tabularnewline
\hline
$8$ & $6$ & $315$ & $3.408095$ & $4.529035$ & $5.075508$ & $5.290515$ & $5.659352$\tabularnewline
\hline
$8$ & $8$ & $405$ & $3.396675$ & $4.529008$ & $5.075504$ & $5.290508$ & $5.659344$\tabularnewline
\hline
$10$ & $8$ & $594$ & $3.278208$ & $4.528883$ & $5.075503$ & $5.290508$ & $5.659332$\tabularnewline
\hline
$10$ & $10$ & $726$ & $3.265916$ & $4.528877$ & $5.075502$ & $5.290507$ & $5.659332$\tabularnewline
\hline
$12$ & $12$ & $1183$ & $3.091015$ & $4.528832$ & $5.075502$ & $5.290506$ & $5.659329$\tabularnewline
\hline
\end{tabular}

\caption{Convergence of total HF energy for a cylindrical potential ($\gamma_{y}=1,\;\gamma_{z}=\sqrt{8}$)
containing two particles, with respect to the size of the basis set,
for various values of the scattering length $a$. Above, $nxmax$
is the maximum value of the quantum number of the SHO basis function
in $x$- and $y-$direction, $nzmax$ is the same number corresponding
to the $z$-direction. Rest of the quantities have the same meaning
as explained in the caption of table \ref{tab:hf-iso-rep}. In some
cases, Fock matrix mixing was employed to achieve convergence. }

\label{tab:hf-cyl}
\end{table}

Inspection of tables \ref{tab:hf-cyl} and \ref{tab:hf-aniso} reveals
that for a given value of interaction length, the convergence requires
the use of larger basis sets with increasing trap anisotropy, ranging
from the perfectly spherical traps, to completely anisotropic traps.
This behavior is expected for cases with aspect ratios $\gamma_{y}$
and $\gamma_{z}>1$, because the effective interaction constant in
such cases $g'=\sqrt{\gamma_{y}\gamma_{z}}g>g$\cite{shukla}.

Upon examining our results for $N=8$ case (\emph{cf.} table \ref{tab:hf-8part}),
we again see very monotonic convergence behavior for all calculations
corresponding to repulsive interactions, and note that the high accuracy
in $E_{HF}$ can be achieved with reasonably sized basis functions.
However, as was the case for $N=2$, completely different behavior
is encountered when the interactions are attractive. The calculations
with $a=-0.05a_{x}$ exhibit systematic convergence in $E_{HF}$ with
the increasing basis set size, but for the case with $a=-0.1a_{x}$,
no trend towards the convergence emerges, pointing again towards an
unstable behavior. 

\begin{table}
\begin{tabular}{|c|c|c|c|c|c|c|c|c|}
\hline 
 &  &  &  & $a=-0.3a_{x}$ & $a=-0.1a_{x}$ & $a=0.1a_{x}$ & $a=0.2a_{x}$ & $a=0.4a_{x}$\tabularnewline
\hline
\hline 
$nxmax$ & $nymax$ & $nzmax$ & $N_{basis}$ & $E_{HF}$ & $E_{HF}$ & $E_{HF}$ & $E_{HF}$ & $E_{HF}$\tabularnewline
\hline
\hline 
$2$ & $0$ & $0$ & $3$ & $4.665163$ & $5.589854$ & $6.372501$ & $6.712470$ & $7.323686$\tabularnewline
\hline 
$2$ & $2$ & $0$ & $9$ & $4.524297$ & $5.577420$ & $6.363976$ & $6.683899$ & $7.236946$\tabularnewline
\hline 
$2$ & $2$ & $2$ & $27$ & $4.377194$ & $5.567347$ & $6.358276$ & $6.665811$ & $7.185985$\tabularnewline
\hline 
$4$ & $2$ & $2$ & $45$ & $4.237680$ & $5.562563$ & $6.357536$ & $6.664895$ & $7.185950$\tabularnewline
\hline 
$4$ & $4$ & $2$ & $75$ & $4.141946$ & $5.560272$ & $6.356979$ & $6.663649$ & $7.184217$\tabularnewline
\hline
$4$ & $4$ & $4$ & $125$ & $4.062610$ & $5.558749$ & $6.356555$ & $6.662580$ & $7.182201$\tabularnewline
\hline
$6$ & $4$ & $4$ & $175$ & $3.914860$ & $5.557403$ & $6.356508$ & $6.662569$ & $7.182068$\tabularnewline
\hline
$6$ & $6$ & $6$ & $343$ & $3.749314$ & $5.556591$ & $6.356406$ & $6.662389$ & $7.181903$\tabularnewline
\hline
$8$ & $8$ & $8$ & $819$ & $3.344656$ & $5.555972$ & $6.356391$ & $6.662373$ & $7.181859$\tabularnewline
\hline
$10$ & $10$ & $10$ & $1331$ & $2.746459$ & $5.555775$ & $6.356390$ & $6.662370$ & $7.181852$\tabularnewline
\hline
$12$ & $12$ & $12$ & $2197$ & $1.871496$ & $5.555707$ & $6.356390$ & $6.662369$ & $7.181851$\tabularnewline
\hline
\end{tabular}

\caption{Convergence of total HF energy for an anisotropic potential ($\gamma_{y}=2,\;\gamma_{z}=3$)
containing two particles, with respect to the size of the basis set,
for various values of the scattering length. Above, $nxmax$, $nymax$,
and $nzmax$ represent the maximum values of the quantum number of
the SHO basis function in $x$-, $y-$, and $z-$ directions, respectively.
Rest of the quantities have the same meaning as explained in the caption
of table \ref{tab:hf-iso-rep}. In some cases, Fock matrix mixing
was employed to achieve convergence.}

\label{tab:hf-aniso}
\end{table}

\begin{table}
\begin{tabular}{|c|c|c|c|c|c|c|c|}
\hline 
 &  & $a=-0.1a_{x}$ & $a=-0.05a_{x}$ & $a=0.01a_{x}$ & $a=0.05a_{x}$ & $a=0.1a_{x}$ & $a=0.2a_{x}$\tabularnewline
\hline
\hline 
$nmax$ & $N_{basis}$ & $E_{HF}$ & $E_{HF}$ & $E_{HF}$ & $E_{HF}$ & $E_{HF}$ & $E_{HF}$\tabularnewline
\hline
\hline 
$1$ & $4$ & $13.05312$ & $15.52656$ & $18.49468$ & $20.47344$ & $22.94688$ & $27.89377$\tabularnewline
\hline 
$3$ & $20$ & $11.11619$ & $15.06420$ & $18.47996$ & $20.16365$ & $21.92218$ & $24.85687$\tabularnewline
\hline 
$5$ & $56$ & $9.55419$ & $14.91302$ & $18.47931$ & $20.16099$ & $21.91766$ & $24.77399$\tabularnewline
\hline 
$7$ & $120$ & $7.88921$ & $14.86390$ & $18.47929$ & $20.15984$ & $21.91127$ & $24.75854$\tabularnewline
\hline 
$9$ & $220$ & $5.87592$ & $14.84991$ & $18.47925$ & $20.15895$ & $21.90931$ & $24.75772$\tabularnewline
\hline
$11$ & $364$ & $3.37633$ & $14.84676$ & $18.47923$ & $20.15863$ & $21.90899$ & $24.75771$\tabularnewline
\hline
$13$ & $560$ & $0.32535$ & $14.84615$ & $18.47921$ & $20.15854$ & $21.90897$ & $24.75767$\tabularnewline
\hline
\end{tabular}

\caption{Convergence of total HF energy ($E_{HF}$) for a spherical symmetric
potential containing eight particles, with respect to the size of
the basis set, for various values of the scattering length. Different
symbols above have the same meaning as explained in the caption of
table \ref{tab:hf-iso-rep}. In all the calculations presented above,
Fock matrix mixing was used to achieve convergence.}

\label{tab:hf-8part}\vspace{0.8cm}

\end{table}

\begin{figure}
\includegraphics[width=8.5cm]{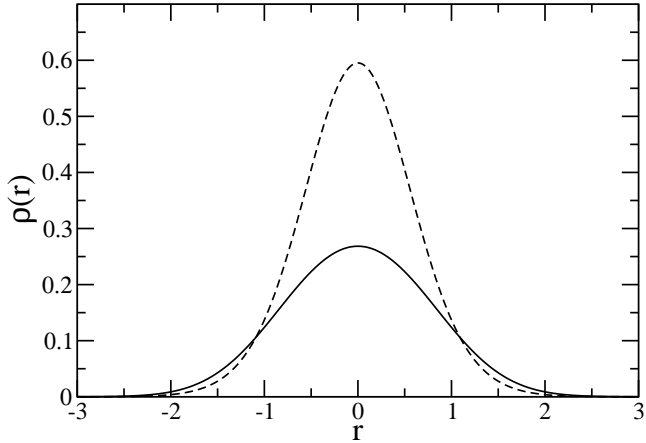}

\caption{Density $\rho(r)=2|\psi_{1s}(r)|^{2}$ plotted along the $x$-axis,
obtained from RHF calculations on a two-particle system in an isotropic
trap with $a=0.2a_{x}$ (solid lines), and $a=-0.2a_{x}$ (dashed
lines). Distance $r$ is in harmonic oscillator units. }
 \label{fig:den-plot}
\end{figure}

Finally, in Fig. \ref{fig:den-plot} we present the orbital density
plots for the $N=2$ case with both attractive and repulsive interactions,
corresponding to $a=\pm0.2a_{x}$. The noteworthy point in the graph
is the accumulation of the density at the center of the trap in case
of attractive interactions, as compared to when the interactions are
repulsive. With increasingly attractive interactions, this phenomenon
becomes even more pronounced, possibly causing the instabilities in
the HF approach.%

\subsection{Unrestricted Hartree-Fock Calculations}

\label{subsec-UHF}

In this section we describe the results of our UHF calculations. If
one performs a UHF calculation on a closed-shell system, one must
get the same results as obtained by an RHF calculation. Similarly,
the total energy and orbitals of a system with $m$ up-spin and $n$
down-spin particles should be the same as that of a system with $n$
up-spin and $m$ down-spin particles. These properties of the UHF
calculations can be used to check the correctness of the underlying
algorithm. We verified these properties explicitly by: (a) performing
UHF calculations on closed-shell systems with various scattering lengths
and geometries, and found that the results always agreed with the
corresponding RHF calculations, and (b) by performing UHF calculations
on various open-shell systems with interchanged spin configurations
and found the results to be identical. Therefore, we are confident
of the essential correctness of our UHF program, and in what follows,
we describe its applications in calculating the addition energy of
fermionic atoms confined in a spherical trap. The aim behind this
calculation is to explore whether such a system follows: (a) shell-structure,
and (b) Hund's rule, in analogy with harmonically trapped electrons
confined in a quantum dot. We also note that a study of Hund's rule
for fermionic atoms confined in an optical lattice was carried out
recently by K\"arkk\"ainen \emph{et al.}\cite{hund-lattice}.

\begin{figure}

\includegraphics[width=12cm]{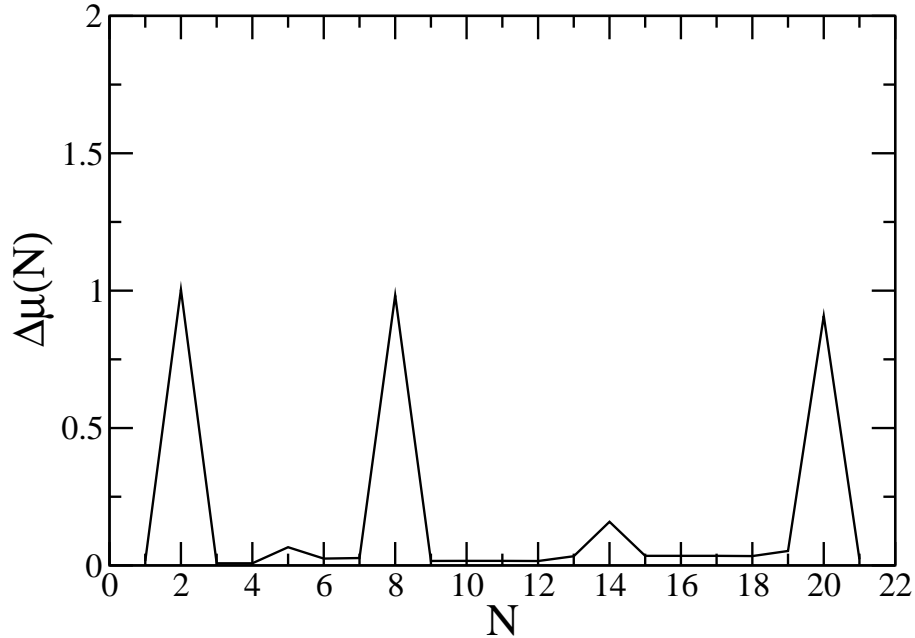}

\caption{Calculated UHF values of addition energies ($\Delta\mu(N)=\mu(N+1)-\mu(N)$)
of a spherical trap (in the units of $\hbar\omega_{x}$) with scattering
length $a=0.01a_{x}$, plotted as a function of the particle number
$N$, ranging from $N=1$ to $N=21$. \label{fig:uhf-delmu}}

\end{figure}

The addition energy, \emph{i.e}, the energy required to add an extra
atom, to an $N$-atom trap is defined as $\Delta\mu(N)=\mu(N+1)-\mu(N)$,
where $\mu(N)$/$\mu(N+1)$ represents the chemical potential of an
$N/(N+1)$ particle system. The chemical potentials, in turn, are
defined as $\mu(N)=E(N)-E(N-1)$, where $E(N)$/$E(N+1)$ represents
the total energy of an $N/(N+1)$ particle system. In our calculations,
the total energies were calculated using the UHF approach for various
values of the scattering length and our results for the addition energy
for an $a=0.01a_{x}$ spherical trap are presented in Fig. \ref{fig:uhf-delmu},
for the values from $N=1$ to $N=21$ .

For the range of $N$ values studied here, in a noninteracting model
the charging energy acquires nonzero values $\Delta\mu(N)=\hbar\omega_{x}$,
only for $N=2,$ $8$, and $20$, corresponding to filled-shell configurations.
In an interacting model, however, $\Delta\mu(N)$ should additionally
exhibit smaller peaks at $N=5$, $N=14$, corresponding to the half-filled
shells. If the inter-particle repulsion is strong enough to split
$3s$ and $3d$ shells significantly, we will additionally obtain
a peak at $N=18$ corresponding to the filled $3d$ shell, while the
peaks corresponding to the half-filled shells will occur at $N=13$,
and $N=19$, instead of $N=14$. Moreover, it is of considerable interest
to examine whether the Hund's rule is also satisfied for open-shell
configurations of such spherically trapped fermionic atoms, as is
the case, \emph{e.g.}, for electrons in quantum dots\cite{hund-qdots}.
From Fig. \ref{fig:uhf-delmu} it is obvious that major peaks are
located at $N=2,$ $8$, and $20$, while the minor ones are at $N=5$,
and $14$, with no peaks at $N=13$, $18$, or $19$. The heights
of the major peaks are in the descending order with increasing $N$,
ranging from $1.003\hbar\omega_{x}$ ($N=2)$ to $0.908\hbar\omega_{x}$
($N=20$). Additionally, for all the open-shell cases, the lowest-energy
configurations were consistent with the Hund's rule in that, a given
shell is first filled with fermions of one (say 'up') spin-orientation,
and upon completion, followed by the fermions of other ('down') spin
orientation. We note that these results are qualitatively similar
to the results obtained for spherical quantum dots\cite{hund-qdots}.
Thus, we conclude that for the small number of particles considered
by us, the shell structure and the Hund's rule are also followed by
atoms confined in harmonic traps where the mutual repulsion is through
short-range the contact interaction. 

We have performed a number of UHF calculations on traps of different
geometries, and scattering lengths, whose results will be published
elsewhere. However, we would like to briefly state that as the scattering
length is increased, in several cases the ferromagnetic configurations
violating the Hund's rule become energetically more stable. This implies
that for large scattering lengths the UHF mean-field approach may
not be representative of the true state, and inclusion of correlation
effects may be necessary.

\section{Conclusions and Future Directions}

\label{sec-conclusions}In this paper we reported a Fortran 90 implementation
of a harmonic oscillator basis set based approach towards obtaining
the numerical solutions of both the restricted, as well as the unrestricted
Hartree-Fock equations for spin-$\frac{1}{2}$ fermions confined by
a harmonic potential, and interacting via pair-wise delta-function
potential. The spin-$\frac{1}{2}$ fermions under consideration could
represent a two-component fermi gas composed of atoms confined in
harmonic traps. We performed a number of calculations assuming both
attractive, and repulsive, inter-particle interactions. As expected,
the Hartree-Fock method becomes unstable with the increasing scattering
length for attractive interactions, while no such problem is encountered
for the repulsive interactions. Additionally, we performed a UHF study
of atoms confined in a spherical harmonic trap and verified the existence
of a shell structure, and that the Hund's rule is followed. These
results are in good qualitative agreement with similar studies performed
on harmonically confined electrons in quantum dots, interacting via
Coulomb interaction. 

In future, we intend to extend and improve the fermionic aspects of
the present computer program in several possible ways. As far as problems
related to fermionic gases in a trap are concerned, we would like
to implement the Hartree-Fock-Bogoliubov approach to allow us to study
such systems in the thermodynamic limit, and at finite temperatures.
With the aim of studying the electronic structure of quantum dots,
we plan to introduce the option of using the Coulomb-repulsion for
interparticle interactions, a step which will require significant
code writing for the two-electron matrix elements. Additionally, we
also aim to introduce the option of studying the dynamics of electrons
in the presence of an external magnetic field, which will also allow
us to study fermionic gases in rotating traps. Finally, we  plan to
implement the option of including spin-orbit coupling in our approach,
which, at present, is a very active area of research. We will report
results along these lines in the future, as and when they become available.


\begin{thebibliography}{10}
\bibitem{fermi-gas-exp}See, \emph{e.g.}, K. M. O'Hara, S. L. Hemmer,
M. E. Gehm, S. R. Granade, and, J. E. Thomas, Science 298 (2002) 2179;
C. A. Regal, C. Ticknor, J. L. Bohn, and D. S. Jin, Nature 424 (2003)
47; M. Greiner, C. A. Regal, and D. S. Jin, 426 (2003) 537; S. Jochim,
M. Bartenstein, A. Altmeyer, G. Hendl, S. Riedl, C. Chin, J. Hecker
Denschlag, and R. Grimm, Science 302 (2003) 2101; M. W. Zwierlin,
C. A. Stan, C. H. Schunk, S. M. F. Raupach, S. Gupta, Z. Hadzibabic,
and W. Ketterle, Phys. Rev. Lett. 91 (2003) 250401; 

\bibitem{new-exp}M. W. Zwierlin, A. Schirotzek, C. H. Schunk, and
W. Ketterle, Science 311 (2006) 492; G. B. Patridge, W. Li, R. I.
Kamar, Y. Liao, R. G. Hulet, Science 311 (2006) 503; M. W. Zwierlin,
C. H. Schunk, A. Schirotzek, and W. Ketterle, Nature 442 (2006) 54. 

\bibitem{fermi-theory-bec-bcs}For a review, see, D. S. Petrov, C.
Salomon, G. V. Shlyapnikov, J. Phys. B 38 (2005) S645; V. Gurarie
and L. Radzihovsky, Ann. Phys. 322 (2002) 2, and references therein.

\bibitem{ogren-shell} Y. Yu, M. \"Ogren, S. \AA berg, S. M. Reimann,
and M. Brack, Phys. Rev. A 72 (2005) 051602(R). 

\bibitem{fermi-gas-theory-phases}See, e.g., P. Pieri and G. C. Strinati,
Phys. Rev. Lett. 96 (2006) 150404; K. B. Gubbels, M. W. J. Romans,
and H. T. C. Stoof, Phys. Rev. Lett. 97 (2006) 210402.t

\bibitem{shukla}R. P. Tiwari and A. Shukla, Comp. Phys. Commun. 174
(2006) 966.

\bibitem{szabo}A. Szabo and N. Ostlund, Modern Quantum Chemistry,
Introduction to Advanced Electronic Structure Theory, Dover Publications,
Inc. (1989). 

\bibitem{lapack}E. Anderson, Z. Bai, C. Bischof, S. Blackford, J.
Demmel, J. Dongarra, J. Du Croz, A. Greenbaum, S. Hammarling, A. McKenney,
and D. Sorensen, LAPACK Users' Guide, 3rd Edn., (2002), SIAM, Philadelphia
(USA).

\bibitem{scatt-length-li6}E. R. I. Abraham, W. I. McAlexander, J.
M. Gerton, R. G. Hulet, R. C\^ot\'e, and A. Dalgarno, Phys. Rev.
A 55 (1997) R3299.

\bibitem{hund-lattice}K. K\"akk\"ainen, M. Borgh, M. Manninen,
and S. M. Reimann, New J. Phys. 9 (2007) 33.

\bibitem{hund-qdots}See, \emph{e.g.}, Y. Asari, K. Takeda, and H.
Tamura, Jpn. J. Appl. Phys. 43 (2004) 4424; C. F. Destefani, J. D.
M. Vianna, and G. E. Marques, arxiv:physics/0404007.
\end{thebibliography}
\end{document}